\documentclass[iop,twocolumn,numberedappendix]{emulateapj}

\usepackage{graphicx}
\usepackage{latexsym}
\usepackage{amsfonts,amsmath,amssymb}
\usepackage{url}
\usepackage[utf8]{inputenc}
\usepackage{fancyref}
\usepackage{hyperref}
\usepackage{natbib}
\usepackage{appendix}
\usepackage{color}

\definecolor{red}{rgb}{1,0,0}
\newcommand{\red}[1]{\textcolor{red}{#1}}
\renewcommand{\red}[1]{#1}

\providecommand{\acknowledgements}{\section{Acknowledgements}}

\slugcomment{To submit to The Astrophysical Journal (ApJ)}

\shorttitle{How Low Can You Go? The Photoeccentric Effect for Planets of Various Sizes}
\shortauthors{Price, Rogers, Johnson, \& Dawson}

\begin{document}

%% LaTeX will automatically break titles if they run longer than
%% one line. However, you may use \\ to force a line break if
%% you desire.

\title{How Low Can You Go? The Photoeccentric Effect for Planets of Various Sizes}

%% Use \author, \affil, and the \and command to format
%% author and affiliation information.
%% Note that \email has replaced the old \authoremail command
%% from AASTeX v4.0. You can use \email to mark an email address
%% anywhere in the paper, not just in the front matter.
%% As in the title, use \\ to force line breaks.

\author{Ellen M. Price}
\affil{California Institute of Technology \\ 1200 East California Boulevard, Pasadena, CA 91125, USA}
\author{Leslie A. Rogers\footnote{Hubble Fellow}}
\affil{Department of Astronomy and Division of Geological and Planetary Sciences \\ California Institute of Technology, MC249-17 \\ 1200 East California Boulevard, Pasadena, CA 91125, USA}
\author{John Asher Johnson\footnote{David \& Lucile Packard Fellow}}
\affil{Harvard-Smithsonian Center for Astrophysics \\ 60 Garden Street, Cambridge, Massachusetts 02138 USA}
\author{Rebekah I. Dawson}
\affil{Department of Astronomy, University of California, Berkeley \\ 501 Campbell Hall \#3411, Berkeley, CA 94720-3411, USA}

\begin{abstract}
It is well-known that the light curve of a transiting planet contains information about the planet's orbital period and size relative to the host star. More recently, it has been demonstrated that a tight constraint on an individual planet's eccentricity can sometimes be derived from the light curve via the ``photoeccentric effect,'' the effect of a planet's eccentricity on the shape and duration of its light curve. This has only been studied for large planets and high signal-to-noise scenarios, raising the question of how well it can be measured for smaller planets or low signal-to-noise cases. We explore the limits of the photoeccentric effect over a wide range of planet parameters. The method hinges upon measuring $g$ directly from the light curve, where $g$ is the ratio of the planet's speed (projected on the plane of the sky) during transit to the speed expected for a circular orbit. We find that when the signal-to-noise in the measurement of $g$ is $< 10$, the ability to measure eccentricity with the photoeccentric effect decreases. We develop a ``rule of thumb'' that for per-point relative photometric uncertainties $\sigma = \{ 10^{-3}, 10^{-4}, 10^{-5} \}$, the critical values of planet-star radius ratio are $R_p / R_\star \approx \{ 0.1, 0.05, 0.03 \}$ \red{for \textit{Kepler}-like 30-minute integration times}. We demonstrate how to predict the best-case uncertainty in eccentricity that can be found with the photoeccentric effect for any light curve. This clears the path to study eccentricities of individual planets of various sizes in the {\emph{Kepler}} sample and future transit surveys.
\end{abstract}

%% Keywords should appear after the \end{abstract} command. The uncommented
%% example has been keyed in ApJ style. See the instructions to authors
%% for the journal to which you are submitting your paper to determine
%% what keyword punctuation is appropriate.

%\keywords{globular clusters: general --- globular clusters: individual(NGC 6397, NGC 6624, NGC 7078, Terzan 8}

\bibliographystyle{apj}

\section{Introduction}
Some planets orbit their stars with fortuitous alignments such that their eclipses can be observed from the Earth. These transiting exoplanets provide a wealth of information about the physical characteristics of planets outside our Solar System. The time interval between successive transit events reveals the orbital period, and the depth of the transit as seen in a photometric time series---the light curve---gives a measure of the planet's radius, assuming that the stellar radius is known \citep{Seager2003,Winn2011,2011exop.book.....S}. In addition to the primary transit event, a secondary eclipse can also be observed when the planet passes behind the star from the observer's vantage point. If this occurs, there is a smaller dip when the light of the planet is blocked by the star, and the depth of the secondary eclipse provides a measure of the planets equilibrium temperature or albedo, depending on the wavelength of observation \citep{RoweEt2008ApJ,2005ApJ...626..523C}. Between eclipse events, phase variations can be observed as different portions of the bright surface of the planet are visible to the observer \citep[e.g.][]{KnutsonEt2007Nature,RoweEt2008ApJ,2010ApJ...723.1436C}

Traditionally, information about a transiting planet's orbit beyond its period, orbital phase, and inclination relative to the sky plane were thought to be the domain of follow-up radial velocity measurements. Specifically, a planet's eccentricity can be readily obtained through time series measurements of the star's reflex motion, in which the planet's eccentricity is manifest as a departure from a purely sinusoidal variation \citep[e.g.][]{2009ApJS..182..205W}. However, highly precise radial velocity measurements require high-resolution spectroscopy, which is expensive in terms of observing time given the faintness of most transiting exoplanetary systems, particularly those discovered by the NASA {\em Kepler} Mission \citep[e.g.][]{BoruckiEt2011ApJ,BatalhaEt2013ApJS}, which have typical magnitudes fainter than $V \approx 12$ \citep{2011AJ....142..112B}. Even using the world's largest telescopes that have precision RV spectrometers such as Keck/HIRES and HARPS-North \citep{2013Natur.503..381H,2013Natur.503..377P}, RV follow-up is only practical for a very small fraction of the more than 3500 {\em Kepler} Objects of Interest.

Fortunately, there is an alternative method of measuring a transiting planet's eccentricity using information encoded in the transit light curve \citep{2007PASP..119..986B, Burke2008ApJ, 2008ApJ...678.1407F, 2012MNRAS.421.1166K, Kipping2014MNRAS}. The eccentricity of a planet's orbit has several observable effects on the transit light curve, and the most notable is a deviation in the duration of a planet's transit compared to an identical planet on a circular orbit of the same period. \red{\citet{WangFord2011} and \citet{MoorheadEt2011} considered this observable in a statistical sample of planets to derive the underlying distribution of eccentricity (see also \citealt{KaneEt2012MNRAS} and \citealt{PlavchanEt2014PASP}). \citet{FQV2008} outlined how the eccentricity could potentially be constrained for individual systems, \red{and \citet{2012MNRAS.421.1166K} used \textit{Multibody Asterodensity Profiling} to constrain the eccentricities of planets in systems with multiple transiting planets}. \citet{DawsonJohnson2012ApJ} recently demonstrated} that the duration and shape deviations, which they coined the ``photoeccentric effect,'' can be used on individual transiting exoplanets using a Bayesian statistical approach to marginalize over the unknown argument of periastron (alignment of the orbit along the line of sight). Their approach takes advantage of the difference in stellar density derived from the transit light curve assuming a circular orbit, $\rho_{\textrm{circ}}$, and the ``true'' stellar density, $\rho_\star$, informed by spectroscopy, stellar \red{isochrons}, and/or asteroseismology. \citet{DawsonJohnson2012ApJ} showed that the photoeccentric effect can effectively measure the eccentricities of highly eccentric, giant planets, even when stellar density is only loosely constrained. Their findings agree well with radial velocity measurements \citep[e.g.][]{DawsonEt2014ApJ}.

\citet{DawsonJohnson2012ApJ} focused on Jupiter-sized planets because such transits have high signal-to-noise and  eccentricities could be verified by subsequent radial velocity measurements. Until now, the question of how well the photoeccentric effect could be used to measure the eccentricities of smaller planets has been left open; for small planets, radial velocity measurements may be expensive or altogether impractical to obtain. Here, we explore the limits of the photoeccentric effect for smaller planets and cases of lower transit signal-to-noise ratio (\red{$\text{SNR}_t$, defined by Equation~\ref{eqn:snr_t}}) using analytic and numerical techniques. In Section~\ref{sec:sigmag}, we introduce our analytic formalism; we go on to discuss the numerical calcuations involved in Section~\ref{sec:sigmae}. In Section~\ref{sec:results}, we discuss our findings. We give examples of applying the photoeccentric effect to planets with low \red{$\text{SNR}_t$} in Section~\ref{sec:examples}. Finally, we discuss the implications of these results in Section~\ref{sec:discussion}.

%Left the question of how well the photoeccentric effect could be measured for smaller planets. Here, we address this question...(give layout of paper).

\section{Methods}
For a planet on a circular orbit with a given orbital period transiting its host star with a given impact parameter, the total transit duration and the timescale of ingress/egress are set by the relative size of the planet's semimajor axis $a$ and the radius of the host star $R_\star$. This is encoded in the scaled semimajor axis, $a/R_\star$, which is a parameter of the transit that can be measured directly from the light curve. Using Newton's version of Kepler's third law, the scaled semimajor axis can be related to the mean stellar density such that $\rho_\star \propto (a/R_\star)^3$ \citep{Seager2003}.

For a planet on an eccentric orbit, the planet will transit its star on a timescale that is typically different than that of a planet with the same period but on a circular orbit. This will yield a transit-derived stellar density that usually differs from the true stellar density. Following \citet{DawsonJohnson2012ApJ}, we define a parameter $g$, which encodes the discrepancy between the stellar density measured from the transit light curve when a circular orbit is assumed, $\rho_{\text{circ}}$, and the ``true'' mean value of stellar density, $\rho_\star$ as

\begin{equation}
\rho_\star = g(e, \omega)^{-3} \rho_{\text{circ}}.
\label{eqn:definition_g}
\end{equation}

\noindent Ultimately, it is the uncertainty in $g$, denoted by $\sigma_g$, that determines the level of confidence with which $e$ can be measured using the photoeccentric effect. The uncertainty in $g$ measured from a high \red{$\text{SNR}_t$} transit around a spectroscopically-characterized star might be estimated on the order of $10\%$ \citep[e.g. as was the case for KOI-1474,][]{DawsonEt2014ApJ}. We expect that $\sigma_g / g$ increases dramatically in lower \red{$\text{SNR}_t$} regimes, in accordance with the increase in the uncertainties of the light curve parameters \citep{PriceRogers2014ApJ}.

Our goal is to quantify how the uncertainty in the eccentricity behaves in different \red{$\text{SNR}_t$} regimes than have been investigated before. We first review how $g$ is related to planet transit parameters, then estimate the precision $\sigma_g$ with which $g$ can be measured in different scenarios, before relating $\sigma_g$ to constraints on the planet orbital eccentricity.

\subsection{Analytic expression for \texorpdfstring{$g$}{g}}
\label{sec:g}
Working from \citet{Kipping2010MNRAS} Equations 30 and 31, and following \citet{DawsonJohnson2012ApJ}, we express the full transit duration (first to fourth contact, \red{$T_{14}$}) and totality duration (second to third contact, \red{$T_{23}$}) as

\begin{align}
T_{14/23} &= \frac{P}{\pi} \frac{\left( 1 - e^2 \right)^{3/2}}{\left( 1 + e \sin \omega \right)^2} \notag\\
&\times \arcsin \left[ \frac{\sqrt{\left( 1 +/- \delta^{1/2} \right)^2 - \left(\frac{a}{R_\star}\right)^2 \left( \frac{1 - e^2}{1 + e \sin \omega} \right)^2 \cos^2 i}}{\left(\frac{a}{R_\star}\right) \left( \frac{1 - e^2}{1 + e \sin \omega} \right) \sin i} \right].
\end{align}

\noindent \red{where $P$ is the orbital period, $e$ is the eccentricity, $\omega$ is the argument of periastron, $i$ is the inclination, $a/R_\star$ is the scaled semimajor axis, and $\delta \equiv (R_p/R_\star)^2$ is the squared scaled planet radius}. Combining the two equations and applying the small angle approximation (which we discuss in Section \ref{sec:approximations}), we can express this formula with the observables on the right-hand side:

\begin{equation}
\frac{a}{R_\star} g(e,\omega) \sin{i} = \frac{2 \delta^{1/4} P}{\pi \sqrt{T_{14}^2 - T_{23}^2}}
\label{eqn:after_smallangle}
\end{equation}

\noindent where

\begin{equation}
g(e,\omega) = \frac{1 + e \sin{\omega}}{\sqrt{1 - e^2}}
\label{eqn:gdefinition}
\end{equation}

Substituting the \citet{DawsonJohnson2012ApJ} Equation 7 definition of $\rho_\text{circ}$, setting $T_{14} = T + \tau$ and $T_{23} = T - \tau$, and approximating $\sin{i} = 1$, we can express $g$ in terms of transit depth $\delta$, transit duration $T$, ingress/egress duration $\tau$, orbital period $P$, and true stellar density $\rho_\star$, as

\begin{equation}
g = \left( \frac{\delta^{1/4}}{\sqrt{T \tau}} \right) \left( \frac{3 P}{G \pi^2 \rho_\star} \right)^{1/3},
\label{eqn:gtrapezoidal}
\end{equation}

\noindent where the $\delta$, $T$, and $\tau$ parameterization of the transit light curve is described in \citet{CarterEtAl2008}.

\subsection{Analytic prediction for \texorpdfstring{$\sigma_g$}{sigma g}}
\label{sec:sigmag}
When the photoeccentric effect is applied in practice, the probability distribution of $g$ for a given planet will be obtained from a numerical fit to the transit light curve (see, e.g., Section~\ref{sec:examples}). This fitting process can be computationally demanding, however. To develop intuition for the behavior of $\sigma_g$ and to explore a wide range of planet scenarios, we estimate the uncertainty on $g$ using a Fisher information analysis and propagation of errors.

To estimate the uncertainty on $g$, we assume that $\delta$, $T$, and $\tau$ are normally-distributed random variables, following the prescription of \citet{CarterEtAl2008}. Furthermore, we assume that $P$ is known to arbitrarily high precision and that $\rho_\star$ is also a normally-distributed random variable. Then, we  analytically determine the variance of $g$ as

\begin{equation}
\sigma_g^2 = \left( \frac{\partial g}{\partial \rho_\star} \right)^2 \sigma_{\rho_\star}^2 + \sum\limits_i \sum\limits_j C_{i,j} \frac{\partial g}{\partial p_i} \frac{\partial g}{\partial p_j}
\label{eq:sigmag}
\end{equation}

\noindent where $C_{i,j}$ is the $(i,j)$ element of the covariance matrix given by Equations 16 and 17 in \citet{PriceRogers2014ApJ} and $\{p\}$ is the set of parameters $\{ t_c, \delta, \tau, T, f_0 \}$, with $t_c$ the time of midtransit and $f_0$ the out-of-transit flux level. Equation~\ref{eq:sigmag} may break down in some regimes, however, specifically at small values of $T$ and $\tau$; we discuss non-Gaussian distributions of $g$ in Section \ref{sec:nongaussian}.

\subsection{Relating \texorpdfstring{$\sigma_g$}{sigma g} to \texorpdfstring{$\sigma_e$}{sigma e}}
\label{sec:sigmae}
\noindent We apply Bayes' theorem to express the the joint posterior distribution of $e$ and $\omega$ conditioned on the available data, $D$, as

\begin{equation}
P\left(e,\omega~|~D\right) \propto \int P\left(D~|~g\right) P\left(g~|~e,\omega\right) P\left(e, \omega\right)~dg.
\end{equation}

\noindent Here, the data, $D$, includes the transit light curve and the observations used to characterize the star. For the purposes of the photoeccentric effect, this data can be distilled into a likelihood function for $g$, $P\left(D~|~g\right)$. We denote the value of $g$ measured from the light curve by $\hat{g}$,  to distinguish it from the unique true value of $g$ for the planet system. We assume, like \citet{DawsonJohnson2012ApJ}, that $\hat{g}$ is a normally distributed random variable with standard deviation $\sigma_g$ centered on the true value of $g$:

\begin{equation}
P\left(D~|~g\right) \equiv P\left(\hat{g}~|~g\right) = \mathcal{N}\left(g,\sigma_g\right)
\end{equation}

\noindent We also express the probability of $g$ conditioned on the eccentricity and argument of periastron,

\begin{equation}
P\left(g~|~e,\omega\right) = \hat{\delta}\left(g-\frac{1 + e \sin{\omega}}{\sqrt{1 - e^2}}\right),
\end{equation}

\noindent where $\hat{\delta}$ is the Dirac delta function. For any $(e,\omega)$ pair, then, we may calculate the likelihood $P(\hat{g}~|~e,\omega)$ using

\begin{eqnarray}
P(\hat{g}~|~e,\omega) &=& \int P\left(\hat{g}~|~g\right) P\left(g~|~e,\omega\right)~dg \\
&=&\frac{1}{\sigma_g \sqrt{2\pi}} \exp{\left( -\frac{\left[ g(e,\omega) - \hat{g} \right]^2}{2\sigma_g^2} \right)}.
\label{eqn:likelihood}
\end{eqnarray}

\noindent The transit probability combined with the condition that the planet's orbit cannot intersect the star describes our prior expectations of $e$ and $\omega$,

\begin{equation}
P(e,\omega) \propto
\begin{cases}
\frac{R_\star}{a} \frac{1 + e \sin{\omega}}{1 - e^2}, & a(1 - e) > R_\star \\
0, & a(1 - e) \le R_\star
\end{cases}.
\label{eqn:prior}
\end{equation}

\noindent We can marginalize the posterior probability, the product of the likelihood and prior probabilities, over $\omega$ to obtain a posterior distribution on eccentricity alone:

\begin{equation}
P(e~|~\hat{g}) = \int P(e,\omega~|~\hat{g})~d\omega \propto \int P(\hat{g}~|~e,\omega) P(e,\omega)~d\omega
\label{eqn:marginalize}
\end{equation}

\noindent We solve this integral numerically to find $\sigma_e$, which we define as half the shortest interval that encloses $68.3\%$ of the area under the curve $P(e~|~\hat{g})$ on its domain $e \in [0,1)$. Note, however, that $e$ will not be normally distributed; we use $\sigma_e$ not as the symmetric width of a normal distribution, but as a way of expressing the confidence interval of $e$ using a widely recognized symbol.

\section{Results}
\label{sec:results}
\begin{figure}
\begin{center}
\includegraphics[width=\columnwidth]{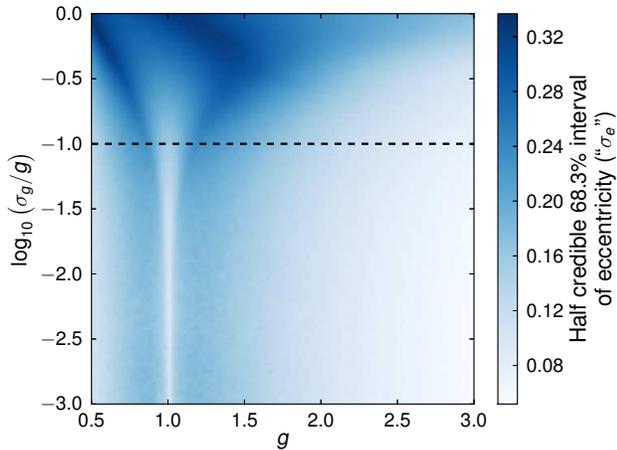}
\caption{\label{fig:contour}
Credible interval of eccentricity $e$ as a function of $g$ and $\sigma_g$, measured numerically by rejection sampling of the posterior pdf, and assuming that $\hat{g}$ is a normally distributed variable: $\hat{g} \sim \mathcal{N}(g, \sigma_g)$. Here we do not include the constraint that the planet cannot intersect the star (Equation \ref{eqn:prior}); including it slightly improves the precision of the eccentricity measurement (smaller $\sigma_e$), particularly for small values of $a/R_\star$. The dashed line indicates $\sigma_g / g = 10\%$; at larger values of the relative uncertainty, the uncertainty in $e$ increases.}
\end{center}
\end{figure}

Under the assumption that $\hat{g}$ is a normally distributed variable with mean $g$ and uncertainty $\sigma_g$, $\sigma_e$ can be estimated directly, because the uncertainties from the light curve parameters are folded into $\sigma_g$. Given $g$ and $\sigma_g$, we calculate a probability for any $(e,\omega)$ pair and then marginalize over $\omega$. In Figure~\ref{fig:contour} we measure the resulting value of $\sigma_e$ as a function of $g$ and the logarithm of its relative uncertainty $\log_{10}{\left( \sigma_g / g \right)}$. We expect that larger $\sigma_g / g$ should result in larger values of $\sigma_e$, and this is what we observe. However, we also notice that values of $g$ far from unity generally result in smaller values of $\sigma_e$ for the same relative uncertainty, so the photoeccentric effect may be applied even in low \red{$\text{SNR}_t$} cases when $e$ is large. We also observe that lower values of $\sigma_e$ can be obtained when $g = 1$; this occurs when $e = 0$ or for appropriate combinations of $e$ and $\omega$. We understand this feature to be a result of the regime transition from $g > 1$ to $g < 1$, at which point the shape of the posterior probability distribution in the $e$, $\omega$ plane changes. Finally, our results suggest a ``rule of thumb'' that, when $\sigma_g / g > 0.1$, or equivalently when the signal-to-noise in the measurement of $g$, \red{$\text{SNR}_g$}, is $< 10$, the ability to measure eccentricity with the photoeccentric effect deteriorates.

\begin{figure}
\begin{center}
\includegraphics[width=\columnwidth]{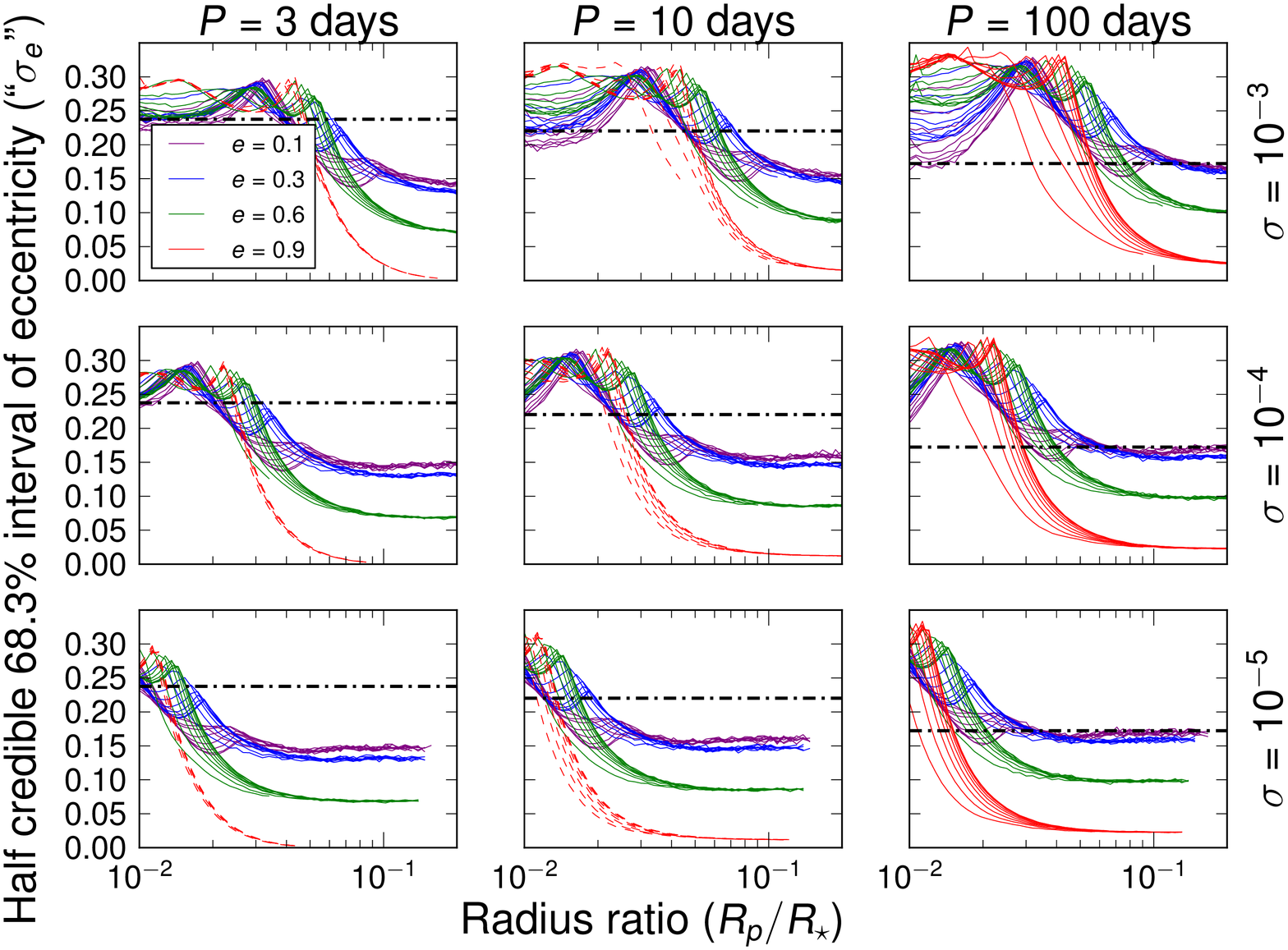}
\caption{\label{fig:spaghetti}
The photoeccentric effect applied to representative cases of orbital period $P$ and \red{per-point relative} photometric uncertainty $\sigma$ for various values of $R_p / R_\star$ \red{and a \textit{Kepler}-like four-year time baseline}; different colors correspond to different values of eccentricity while individual lines represent different values of the impact parameter $b$. \red{We assume a precisely-known stellar density such that $\sigma_{\rho_\star} = 0$}. For low photometric precision and small $R_p / R_\star$, the eccentricity posterior is prior-dominated, which results in a measurement of \red{moderate precision but low accuracy}. As the posterior becomes dominated by the likelihood, the uncertainty increases slightly and then decreases as the prior has less influence on the posterior. Dashed lines indicate that $\left(a / R_\star\right)^2 < \frac{2}{3} \left( 1 + e \right)^3 / \left( 1 - e \right)^3$, in which case the condition of \citet{Kipping2014MNRAS} Equation B14 certainly fails (see Section \ref{sec:approximations} for a discussion of the approximations that can break down in this analysis). We also plot the width of the prior distribution (marginalized over $\omega$), the value to which we expect the uncertainty in $e$ to asymptote in the limit of completely uninformative data, in dash-dot lines. See Appendices \ref{adx:small} and \ref{adx:large} for discussions of the small and large $R_p/R_\star$ limits.}
\end{center}
\end{figure}

We have found that the assumption $g(e, \omega) \sim \mathcal{N}\left( g, \sigma_g \right)$ can break down at small \red{$\text{SNR}_t$} (see Section~\ref{sec:nongaussian}), so we give several representative, idealized measurements of $\sigma_e$ in Figure~\ref{fig:spaghetti} by calculating the distribution of $g$ numerically and parametrizing in terms of variables for which astronomers have better intuition. We assume that $\delta$, $T$, and $\tau$ are normally distributed with the variances and covariances predicted for binned light curves by \citet{PriceRogers2014ApJ}, \red{and we use Equation~\ref{eqn:gtrapezoidal} to calculate a distribution of $g$ from these distributions}. \red{We also assume the orbital period $P$ and stellar density $\rho_\star = \rho_\odot$ are known to absolute precision for simplicity, making this prediction a lower bound on the uncertainty in $e$.} Again, $e$ is better constrained when eccentricity is large. In all cases, there is a ``critical'' value of $R_p / R_\star$ below which $\sigma_e$ sharply increases, and the critical value is a function of all the transit parameters.

\begin{figure}
\begin{center}
\includegraphics[width=\columnwidth]{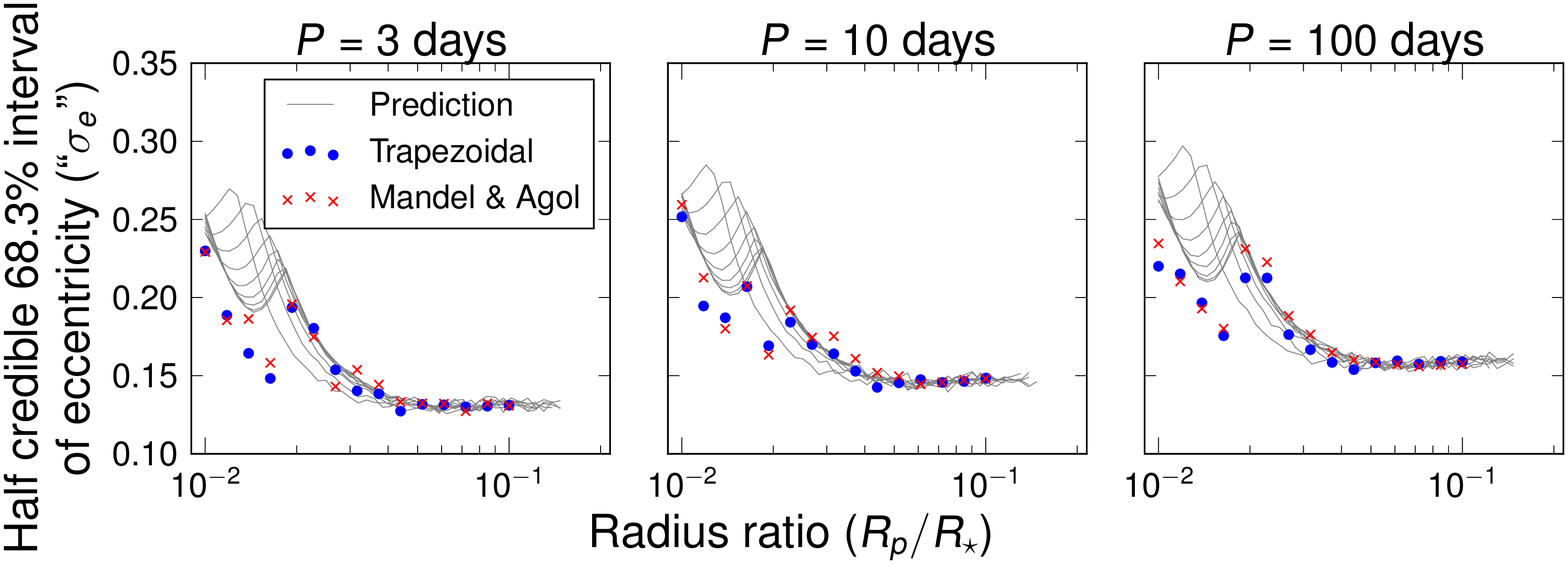}
\caption{\label{fig:mcmc}
We test the validity of the numerical predictions shown in Figure~\ref{fig:spaghetti}, which are shown here as gray lines, by performing MCMC fits to synthetic transit light curves. We assume a \red{relative photometric uncertainty of $\sigma = 10^{-5}$ on each $30$-minute integrated time point}, which corresponds to the bottom row of panels in Figure \ref{fig:spaghetti}; individual lines indicate different values of the impact parameter. In our synthetic light curves, we choose $b = 0.1$, $\omega = \pi / 2$, \red{and $e = 0.3$}. Fitting with a \citet{CarterEtAl2008} trapezoidal model (blue points) and a \citet{MandelAgol2002ApJ} quadratically limb-darkened model (red crosses) yield similar results, even though the predictions are based on a trapezoidal model.}
\end{center}
\end{figure}

We perform numerical experiments estimating the posterior from synthetic light curves using Markov chain Monte Carlo to support our predictions. \red{We fit synthetic, \citet{MandelAgol2002ApJ} light curves} \red{on a \textit{Kepler}-like four-year time baseline} with both the \citet{CarterEtAl2008} trapezoidal model and the \red{\citeauthor{MandelAgol2002ApJ} quadratically limb-darkened model}, for which we use a Python adaptation of the \citet{EastmanEt2013PASP} EXOFAST code. Our fitting procedure uses the Python \texttt{emcee} module's affine-invariant ensemble sampler \citep[][proposed by \citealp{GoodmanWeare2010}]{ForemanMackey2013PASP}, resulting in $3 \times 10^4$ posterior distribution samples. \red{In the case of the \citeauthor{MandelAgol2002ApJ} fit, we fit in terms of the \citeauthor{CarterEtAl2008} trapezoidal parameters, transforming to the physical parameters to evaluate the model function, because they are less correlated than physically-motivated parameters like $a/R_\star$.} We assume a \red{relative photometric uncertainty of $\sigma = 10^{-5}$ on each $30$-minute integrated time point}, eccentricity $e = 0.3$, argument of periastron $\omega = \pi / 2$, impact parameter $b = 0.1$, \red{and stellar density uncertainty $\sigma_{\rho_\star} = 0$} for the purposes of this test. We find that our predictions are valid for both models (see Figure~\ref{fig:mcmc}).

\section{Example Applications}
\label{sec:examples}
We now turn to applying the photoeccentric effect to measure the eccentricity of known transiting planets from their transit light curves. Our aims are both to test the semi-analytic estimates of $\sigma_e$ by comparing the predictions to numerically determined credible intervals for $e$, and to test the accuracy of the photo-eccentricity constraints in the low \red{$\text{SNR}_t$} regime by comparing the photo-eccentricities to the RV-measured $e$ values. These examples also serve to highlight the power and limitations of photo-eccentricities.

Given an arbitrary transit light curve, we use forward modelling to generate the joint distribution of $e$ and $\omega$. We use a Python adaptation of the \citet{EastmanEt2013PASP} implementation of the \citet{MandelAgol2002ApJ} limb-darkened light curve model to fit the light curve data in terms of the \citet{CarterEtAl2008} trapezoidal shape parameters (which are less correlated than physically-motivated parameters like $a/R_\star$ \red{but which we transform to the physical parameters to evaluate the model function}) and two limb darkening parameters $q_1$ and $q_2$ \citep{Kipping2013MNRAS}, which transform to the \citet{MandelAgol2002ApJ} parameters $u_1$ and $u_2$. We use the Python \texttt{emcee} module \citep{ForemanMackey2013PASP} with $3 \times 10^5$ MCMC chain samples to perform these fits. For each set of parameters in the chain, we calculate an estimate of stellar density,

\begin{equation}
\rho_\text{circ} = \frac{M_\star + M_p}{\frac{4}{3} \pi R_\star^3} \approx \frac{M_\star}{\frac{4}{3} \pi R_\star^3} = \frac{3 \pi}{G P^2} \left( \frac{a}{R_\star} \right)^3,
\end{equation}

\noindent which follows directly from Newton's version of Kepler's third law (assuming a circular orbit and $M_p \ll M_\star$). The parameter $g$ can be found from Equation \ref{eqn:definition_g} \red{by drawing normally-distributed random samples from $\mathcal{N} \left( \rho_\star, \sigma_{\rho_\star} \right)$, where $\sigma_{\rho_\star}$ is set by the independent observational constraints on $\rho_\star$}.

\red{We perform a second MCMC exploration in $(e,\omega,\rho_\star)$ parameter space, using the observed distribution of $\rho_\text{circ}$ from photometry and the observed distribution of $\rho_\star$ from the literature; we do not fit the light curve directly at this step but instead use the posteriors from the circular fit. This yields posterior distributions of $e$ and $\omega$ consistent with the parameters measured assuming a circular orbit. This procedure is advantageous because it allows us to fit eccentricity separately from the light curve shape parameters; fitting the shape parameters, $e$, and $\omega$ together is computationally intensive. This step is also necessary because the periapse distance constraint, Equation~\ref{eqn:prior}, depends on $a/R_\star$, which is not held fixed as in Section~\ref{sec:results}; instead, it is a distribution, determined by the distribution of $\rho_\star$.} Marginalization over the $\omega$ nuisance parameter, \red{via Equation~\ref{eqn:marginalize}}, and \red{marginalization over $\rho_\star$} allows us to solve for the credible interval of $e$ numerically.

\subsection{HAT-P-2b = \texorpdfstring{HD\,147506}{HD 147506}}
\label{sec:hatp2}
\begin{figure}
\begin{center}
\includegraphics[width=\columnwidth]{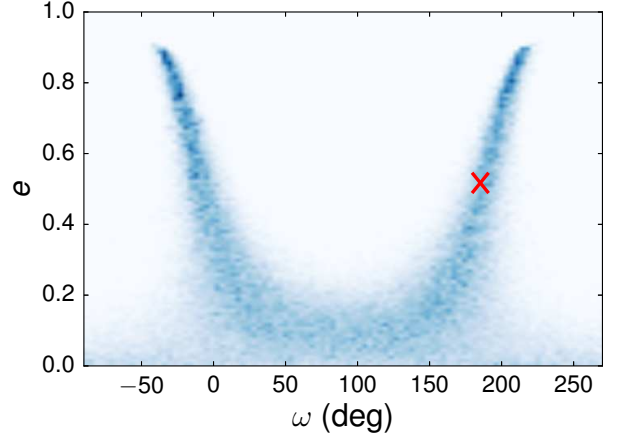}
\caption{\label{fig:hatp2-2d}
Two-dimensional posterior probability distribution $P(e, \omega)$ for HAT-P-2b. The ``true'' values of $e$ and $\omega$ measured by \citet{PalEt2010ApJ}, indicated by a red cross, are allowed by this distribution with nonzero probability. The high probabilities concentrated at large $e$ are results of the prior, the probability of nongrazing transit.}
\end{center}
\end{figure}

\begin{figure}
\begin{center}
\includegraphics[width=\columnwidth]{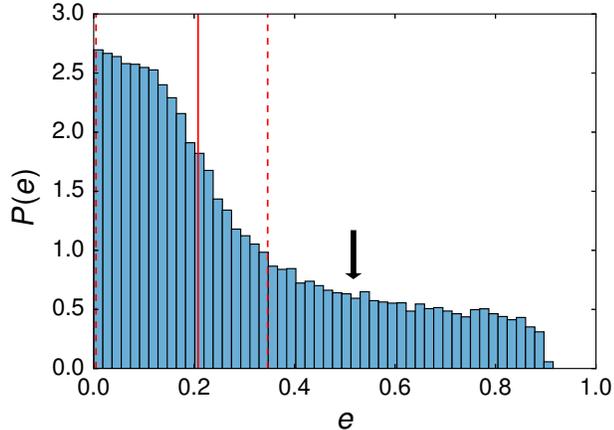}
\caption{\label{fig:hatp2-1d}
Posterior probability distribution $P(e)$ for HAT-P-2b. The $68.3\%$ credible interval (red dashed lines), indicated around the \red{median} (solid red line), does not enclose the measured value (indicated by an arrow), but the posterior does not disallow high values of eccentricity.}
\end{center}
\end{figure}

We fit the phase-folded photometry data of HAT-P-2b \red{($P = 5.63~\textrm{days}$)} from \citet{PalEt2010ApJ} with the model described above. We measure the \citet{CarterEtAl2008} trapezoidal shape parameters $\delta = 0.0052 \substack{+0.00009 \\ -0.00009}$, $T = 0.164 \substack{+0.0007 \\ -0.0008}$ days, and $\tau = 0.0129 \substack{+0.0009 \\ -0.0009}$ days; we estimate the signal-to-noise ratio to be $\delta / \sigma_\delta \approx 59$. We adopt $M_\star = 1.308 \substack{+0.088 \\ -0.078}~M_\odot$ and $R_\star = 1.506 \substack{+0.13 \\ -0.096}~R_\odot$ from \citet{TorresWinnHolman2008ApJ}, derived from stellar evolution models. From those values, we estimate $\rho_\star = 0.56 \pm 0.14~\rm{g\,cm^{-3}}$. This culminates in an estimate of \red{$e = 0.21 \substack{+0.14 \\ -0.20}$, the $68.3\%$ confidence interval around the median of the distribution}. The measurement used for Figure \ref{fig:spaghetti} predicts the value of ``$\sigma_e$'' to be about 0.151, and we measure \red{``$\sigma_e$'' $= 0.171$ from the MCMC posterior}. The full two-dimensional posterior probability distribution $P(e, \omega)$ is shown in Figure~\ref{fig:hatp2-2d}, and the marginalized posterior probability $P(e)$ is shown in Figure \ref{fig:hatp2-1d}.

The HAT-P-2b system exemplifies a case for which the light curve is relatively uninformative about the eccentricity. \citet{PalEt2010ApJ} measured the eccentricity of HAT-P-2b to be $e = 0.5171 \pm 0.0033$; this RV $e$ measurement falls outside the 68.3\% credible interval for $e$ derived from the photoeccentric effect. Examining the two-dimensional posterior probability distribution in Figure \ref{fig:hatp2-2d} reveals that there is nonzero probability of the true value of eccentricity for $\omega \approx \pi$, and \citet{PalEt2010ApJ} measured $\omega = 185.22^{\circ} \pm 0.95^{\circ}$ for this planet. \red{Although outside the 1--$\sigma$ confidence region, the true value for $e$ lies within the statistically allowed constraints of the marginalized posterior distribution for $e$ derived from our analysis.}

\subsection{GJ-436b}
\label{sec:gj436}
\begin{figure}
\begin{center}
\includegraphics[width=\columnwidth]{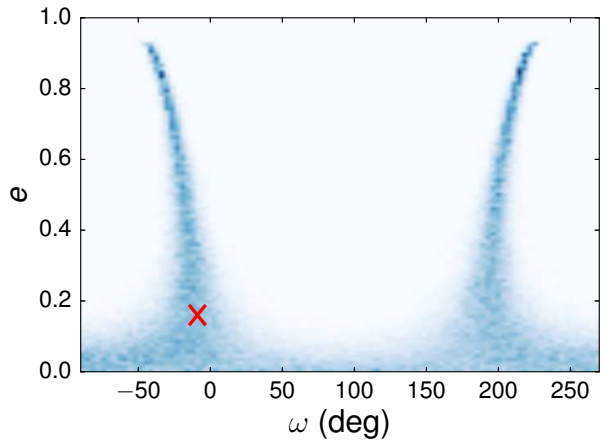}
\caption{\label{fig:gj436-2d}
Two-dimensional posterior probability distribution $P(e, \omega)$ for GJ~436b. The values for $e$ and $\omega$ measured by \citet{ManessEt2007PASP}, indicated by a red cross, are allowed by this posterior. As seen before in the case of HAT-P-2b, the high probabilities concentrated at high $e$ are the result of the prior imposed.}
\end{center}
\end{figure}

\begin{figure}
\begin{center}
\includegraphics[width=\columnwidth]{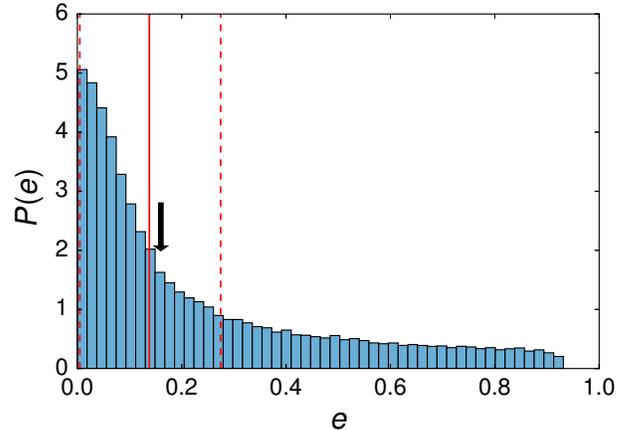}
\caption{\label{fig:gj436-1d}
Posterior probability distribution $P(e)$ for GJ~436b. The $68.3\%$ credible interval (red dashed lines), indicated around the \red{median} (solid red line), encloses the measured value, indicated by an arrow.}
\end{center}
\end{figure}

In our second application of the photoeccentric effect, we focus on GJ~436b \red{($P = 2.64~\textrm{days}$, \citealt{vonBraunEt2012ApJ})}, a Neptune-size planet on an eccentric orbit around an M dwarf star. We fit the phase-folded photometry data for the transit of GJ~436b, as observed by the \textit{Spitzer} Space Telescope on 2 February 2009 \citep[see][]{KnutsonEt2011ApJ}, with the \citet{MandelAgol2002ApJ} model. We measure the \citet{CarterEtAl2008} trapezoidal shape parameters to be $\delta = 0.0071 \substack{+0.0002 \\ -0.0002}$, $T = 0.0333 \substack{+0.0006 \\ -0.0004}$ days, and $\tau = 0.0102 \substack{+0.0007 \\ -0.0007}$ days, so we estimate the signal-to-noise ratio as $\delta / \sigma_\delta \approx 30$. We use an interferometric radius measurement for the host star of GJ~436b from \citet{vonBraunEt2012ApJ}, which gives $R_\star = 0.455 \pm 0.018~R_\odot$. We also obtain a mass estimate from \citet{Torres2007ApJL}, which gives $M_\star = 0.452 \substack{+0.014 \\ -0.012}~ M_\odot$, from $J - K$ and $M_K$. From these values, we estimate a stellar density of $\rho_\star = 6.83 \pm 0.846~\rm{g\,cm^{-3}}$. We measure \red{$e = 0.14 \substack{+0.14 \\ -0.13}$ around the median of the distribution}. Thus our ``$\sigma_e$'' \red{measured from the MCMC posterior} is about 0.136, and the numerical measurement we use in for Figure \ref{fig:spaghetti} predicts 0.134. The full two-dimensional posterior probability distribution $P(e, \omega)$ is shown in Figure~\ref{fig:gj436-2d}, and the marginalized posterior probability distribution $P(e)$ is shown in Figure \ref{fig:gj436-1d}.

\citet{ManessEt2007PASP} measured the eccentricity and argument of periastron of GJ~436b to be $e = 0.160 \pm 0.019$ and $\omega = 351^{\circ} \pm 1.2^{\circ}$. From the two-dimensional posterior in Figure \ref{fig:gj436-2d}, the ``true'' values have nonzero probability. Furthermore, we are able to recover $e$ to within the credible interval after marginalizing over $\omega$, unlike the case of HAT-P-2b.

\section{Discussion}
\label{sec:discussion}

\subsection{Scaling with signal-to-noise}
From Figure \ref{fig:spaghetti}, we see that there is a rapid increase in $\sigma_e$ as $R_p/R_\star$ decreases; past this threshold, eccentricity is constrained very poorly. We use the \citet{GaudiEt2004ApJ} definition of total transit signal-to-noise,

\begin{equation}
\red{\text{SNR}_t = Q = N_p^{1/2} \left( \frac{\delta}{\sigma} \right),}
\label{eqn:snr_t}
\end{equation}

\noindent where $N_p$ is the total number of measurements during transit, $\delta$ is the transit depth, and $\sigma$ is the per-point uncertainty. Among the cases shown in Figure \ref{fig:spaghetti}, the upturn in $\sigma_e$ occurs at a total \red{$\text{SNR}_t$} between about $100$ and $1000$. To gain useful insights into a planet's eccentricity from a transit light curve alone, therefore, a high \red{$\text{SNR}_t$} is needed. Planets included in the KOI catalog will have a minimum $\text{SNR}_t$ of 7.1 \citep{BatalhaEt2010ApJL,BoruckiEt2011ApJ}, but useful eccentricity measurements will generally require much higher \red{$\text{SNR}_t$}.

\red{The signal-to-noise in $g$, $\text{SNR}_g$, is an increasing function of $Q$, but, due to the covariances between the transit observables, it does not only depend on $Q$ but also on the precise combination of orbital properties. Since $\sigma_e$ must be measured numerically from the distribution of $g$, it is an even more complicated function of $Q$. As a result, the dimension of the of the grid in Figure~\ref{fig:spaghetti} is not reduced when recast in terms of $Q$, and the location of the ``knee'' in that figure spans a range of $Q$ values.}

The transit signal-to-noise at which the threshold occurs depends on $\sigma$ and other transit parameters, but this is to be expected. The estimate that we use for the \red{$\text{SNR}_t$} does not take into account the effects of finite exposure time, which has a stronger effect on $\sigma_e$ when $R_p/R_\star$ is small because of shorter ingress/egress times \citep{PriceRogers2014ApJ}. The total \red{$\text{SNR}_t$} of the detected transit also does not account for the number of photometric points taken during ingress and egress, but rather the entire transit; measuring $\tau$ helps reduce the degeneracies between impact parameter, transit duration, and $R_p/R_\star$.

\subsection{Non-Gaussian distributions of \texorpdfstring{$g$}{g}}
\label{sec:nongaussian}
In regimes of low signal-to-noise, our approximation that $g$ is a normally distributed variable (see Equation \ref{eqn:likelihood} \red{and Figure~\ref{fig:contour}}) may break down. Using the estimates of $\sigma_\delta$, $\sigma_T$, and $\sigma_\tau$ from \citet{PriceRogers2014ApJ}, we see that small values of $R_p / R_\star$ can yield distributions of $T$ and $\tau$ (transit duration and ingress/egress duration, respectively, from the trapezoidal transit model) which include and are truncated at $0$, since negative durations would be unphysical. When we calculate $g$ using Equation \ref{eqn:gtrapezoidal} with distributions of $\delta$, $T$, and $\tau$, the resulting distribution on $g$ resembles a log-normal distribution because of the vanishingly small denominator for some $T$ and $\tau$.

We encountered non-Gaussian $g$ distributions at small $R_p/R_\star$ in Figure~\ref{fig:spaghetti} and in the example applications (Section~\ref{sec:examples}). The solution we developed was to calculate the distribution of $g$ non-parametrically (by linearly interpolating the pdf) instead of making the normally-distributed variable assumption.

\subsection{Breakdown on miscellaneous approximations}
\label{sec:approximations}
In the derivations of Equation \ref{eqn:after_smallangle}, we made the same approximation as \citet{DawsonJohnson2012ApJ} and \citet{Kipping2010MNRAS} in asserting that the quantity

\begin{equation}
\frac{\sqrt{\left( 1 + / - \delta^{1/2} \right)^2 - \left(\frac{a \left( 1 - e \right) }{R_\star}\right)^2 \left( \frac{1 + e}{1 + e \sin \omega} \right)^2 \cos^2 i}}{\left(\frac{a \left( 1 - e \right)}{R_\star}\right) \left( \frac{1 + e}{1 + e \sin \omega} \right) \sin i}
\end{equation}

\noindent is small, \red{which follows from the assumption that the \textit{arcsin} term itself is small}. \red{This assumption is violated when $\sin i$ is small, which also invalidates the assumption made to obtain Equation~\ref{eqn:gtrapezoidal}, the definition of $g$ in terms of the light curve shape parameters; to obtain that result, we assumed $\sin i \approx 1$, which is true unless the planet both comes within just a few stellar radii of its host during transit and has a large impact parameter. When this approximation breaks down, i.e. when the separation during transit divided by the stellar radius approaches unity, the photoeccentric effect as it is presented here will not apply.} \citet{Kipping2014MNRAS} derives the conservative condition

\begin{equation}
\left( a / R_\star \right)^2 \gg \frac{2}{3} \frac{\left( 1 + e \right)^3}{\left( 1 - e \right)^3}
\end{equation}

\noindent (his Equation 35) under which the sine small-angle approximation and inverse sine small-angle approximation should be valid. This condition should be checked, particularly for systems with small orbital periods and large eccentricities.

By parameterizing Equation \ref{eqn:gtrapezoidal} in terms of the \citet{CarterEtAl2008} trapezoidal light curve parameters, we have implicitly assumed a symmetric transit shape. While this parameterization is suitable for the error analysis in Section \ref{sec:sigmag}, we could more accurately use $T_{14}$ and $T_{23}$ as the times between first and fourth contacts and second and third contacts, respectively.

Since the parameter $g$ is the ratio of the planet's velocity during transit to the velocity assuming $e = 0$, it is necessary to approximate the in-transit velocity as being constant across the stellar disk. We note that this approximation breaks down for planets with long transit durations (compared to the orbital period).

Finally, both the error analysis of \citet{PriceRogers2014ApJ} and the Markov chain Monte Carlo fits we have performed assume flat priors on the trapezoidal light curve parameters. We have held to this assumption for self-consistency. Assuming different priors on these parameters or flat priors on more physically-motivated parameters would change the expected value of $\sigma_e$, for example, given a particular set of orbit parameters. This effect should be most important in a prior-dominated regime, however, and the location of the increase in $R_p / R_\star$ should be relatively insensitive to the prior used.

\red{For a discussion of the effects of blending, spots, and TTVs on measuring eccentricity, we refer the reader to \citet{Kipping2014MNRAS}.}

\section{Summary}
\label{sec:summary}
We present here analytic and numeric approximations for how well the photoeccentric effect may be applied in various signal-to-noise regimes, taking into account other transit parameters, such as orbital period. The method we present generally works best for very small and very large eccentricities; intermediate values of eccentricity often result in wide posterior probability distributions that do not allow eccentricity to be constrained as well. \red{When the signal-to-noise in the measurement of $g$, \red{$\text{SNR}_g$}, is $< 10$, the ability to measure eccentricity with the photoeccentric effect decreases significantly.} The uncertainty on eccentricity increases monotonically with decreasing \red{transit} signal-to-noise, \red{$\text{SNR}_t$}, until a critical value, at which the posterior becomes uninformative. This value depends on multiple orbital parameters, including the orbital period and impact parameter, in addition to the signal-to-noise ratio, as shown in Figure \ref{fig:spaghetti}. Based on this figure, we developed a ``rule of thumb'' that for per-point relative photometric uncertainties $\sigma = \{ 10^{-3}, 10^{-4}, 10^{-5} \}$, the critical values of planet-star radius ratio are $R_p / R_\star \approx \{ 0.1, 0.05, 0.03 \}$ \red{for \textit{Kepler}-like 30-minute integration times}.

\acknowledgements
We would like to thank A. P{\'{a}}l and H. Knutson for use of their photometry data. EMP acknowledges funding provided by Shirley and Carl Larson for her 2013 Carolyn Ash SURF Fellowship. LAR acknowledges support provided by NASA through Hubble Fellowship grant \#HF-51313.01 awarded by the Space Telescope Science Institute, which is operated by the Association of Universities for Research in Astronomy, Inc., for NASA, under contract NAS 5-26555. JAJ is grateful for the generous grant support provided by the Alfred P. Sloan and David \& Lucile Packard foundations. RID gratefully acknowledges support by the Miller Institute for Basic Research in Science at University of California, Berkeley.

\begin{appendix}
\newcommand{\sigmae}{\sigma_e}
\newcommand{\rprstar}{R_p / R_\star}

\section{Asymptotic behavior of \texorpdfstring{$\protect\sigmae$}{sigma e} for small \texorpdfstring{$\protect\rprstar$}{Rp/R*}}
\label{adx:small}
In the limit of small $R_p / R_\star$, we assume that $g$ is so poorly constrained that the distribution of $e$ reduces to that of the prior. That is, $P(D~|~e) \propto 1$, so

\begin{equation}
P(e~|~D) \propto P(D~|~e) P(e) \propto P(e)
\end{equation}

\noindent and the posterior distribution becomes, if the transit probability is imposed as a prior,

\begin{equation}
P(e~|~D) \propto
\begin{cases}
\frac{1 + e \sin{\omega}}{1 - e^2}, & a(1 - e) > R_\star \\
0, & a(1 - e) \le R_\star
\end{cases}.
\end{equation}

\noindent The uncertainty in $e$ is just the measurement of the width of the prior, which must be done numerically. The value of the uncertainty in this regime depends only on the scaled semimajor axis $a/R_\star$, which sets the maximum allowed value of $e$.

\section{Asymptotic behavior of \texorpdfstring{$\protect\sigmae$}{sigma e} for large \texorpdfstring{$\protect\rprstar$}{Rp/R*}}
\label{adx:large}
We now turn to explaining the asymptotic behavior of $\sigma_e$ in the limit of large $R_p / R_\star$. We begin with Bayes' theorem, to write the posterior probability of $e$ as

\begin{equation}
P\left(e~|~D\right) = \int P\left(e,\omega~|~D\right)~d\omega \propto \int \int P\left(D~|~g\right) P\left(g~|~e,\omega\right) P\left(e, \omega\right)~dg~d\omega.
\end{equation}

\noindent When the \red{$\text{SNR}_g$} ratio is large (i.e. when $\sigma_g \rightarrow 0$), we approximate

\begin{equation}
P\left(D~|~g\right) = \mathcal{N}\left(g,\sigma_g\right) \approx \hat{\delta}\left(\hat{g} - g\right),
\end{equation}

\noindent where $\hat{g}$ is the value of $g$ measured from the transit data, and $\hat{\delta}$ is the Dirac delta function. We also express

\begin{equation}
P\left(g~|~e,\omega\right) = \hat{\delta}\left(g-\frac{1 + e \sin{\omega}}{\sqrt{1 - e^2}}\right),
\end{equation}

\noindent to obtain

\begin{eqnarray}
P\left(e~|~D\right) &\propto&  \int \int \hat{\delta}\left(\hat{g} - g\right) \hat{\delta}\left(g-\frac{1 + e \sin{\omega}}{\sqrt{1 - e^2}}\right) P\left(e, \omega\right)~dg~d\omega\\
&=& \int \hat{\delta}\left(\hat{g}-\frac{1 + e \sin{\omega}}{\sqrt{1 - e^2}}\right) P\left(e, \omega\right)~d\omega.
\end{eqnarray}

\noindent At this point in the proof, we use the composition rule for $\hat{\delta}$ functions. The argument of the $\hat{\delta}$ has simple zeroes at $\omega_1 = \sin^{-1}\left( {\frac{\hat{g}\sqrt{1 - e^2} - 1}{e}} \right)$ and $\omega_2 = \pi - \sin^{-1}\left( {\frac{\hat{g}\sqrt{1 - e^2} - 1}{e}} \right)$, so we can write it as

\begin{align}
\hat{\delta}\left(f\left(\omega\right)\right) &= \sum\limits_i \frac{\hat{\delta}\left(\omega - \omega_i\right)}{\left| f'\left(\omega_i\right) \right|} \\
&= \sum\limits_i \hat{\delta} \left(\omega - \omega_i\right) \left[ \frac{2 \hat{g}}{\sqrt{1 - e^2}} -\left(1 + \hat{g}^2\right)\right]^{-1/2}.
\end{align}

\noindent The posterior pdf of eccentricity is then

\begin{equation}
P\left(e~|~D\right) \propto \int \sum\limits_{i} \hat{\delta} \left(\omega - \hat{\omega}_i\right) \left[ \frac{2 \hat{g}}{\sqrt{1 - e^2}}-\left(1 + \hat{g}^2\right) \right]^{-1/2} P\left(e, \omega\right)~d\omega.
\end{equation}

\noindent Assuming flat priors on $e$ and $\omega$, $P\left(e, \omega\right) \propto 1$, the pdf becomes

\begin{equation}
P(e~|~D) \propto \left[ \frac{2 \hat{g}}{\sqrt{1 - e^2}}-\left(1 + \hat{g}^2\right) \right]^{-1/2}.
\end{equation}

\noindent We could alternatively use the transit probability as a prior on $e$ and $\omega$,

\begin{equation}
P(e, \omega) \propto \frac{1 + e \sin{\omega}}{1 - e^2} \propto \frac{g}{\sqrt{1 - e^2}},
\end{equation}

\noindent in which case the pdf is

\begin{equation}
P(e~|~D) \propto \left( \frac{\hat{g}}{\sqrt{1 - e^2}} \right) \left[ \frac{2 \hat{g}}{\sqrt{1 - e^2}} -\left(1 + \hat{g}^2\right)\right]^{-1/2}.
\end{equation}

\noindent These analytic pdfs qualitatively agree with a numerically integrated joint $(e,\omega)$ posterior as $\sigma_g \rightarrow 0$.

A functional form of both cdfs can be calculated as well, to normalize $P(e~|~D)$ and to calculate $\sigma_e$. The cdfs contain elliptic integrals, however, which make them less informative for building intuition from analytic expressions yet computationally favorable for calculating the integrals numerically.
\end{appendix}

\vspace{1em}
\bibliography{biblio}

\end{document}